\documentstyle[11pt,paspconf,epsf]{article}

\def \hii {H\,II}
\def \hi {H\,I}

\begin{document}

\title{Star Formation and Abundances in S0 Galaxies}

\author{Richard W. Pogge \& Paul B. Eskridge}
\affil{Astronomy Department, The Ohio State University,
    Columbus, OH, 43210-1106}

\begin{abstract}
A significant number of S0 galaxies with a detectable ISM show some level
of on-going massive-star formation activity in the form of visible \hii\
Regions.  A rich ISM, however, does not guarantee star formation: a
significant number of relatively gas-rich S0s have no detectable \hii\
regions down to the level of regions ionized by single O stars (e.g.,
Orion-like \hii\ regions).  The H$\alpha$ luminosities of star-forming S0s
imply global star formation rates as much as 2--3 orders of magnitude lower
than in spirals.  Spectroscopy of the bright \hii\ regions in a few of the
star-forming S0s studied thus far reveal solar gas-phase oxygen abundances.
In general, S0s fall onto the general trends of O/H with galaxy luminosity
and Hubble Type seen for Spirals and Irregulars, and have O/H abundances
typical of spirals of similar total stellar luminosity.  The implications
are that S0s appear to be the low end of the Hubble Sequence of spirals,
and that their interstellar gas is primarily internal in origin.  The
surprise is that two counter-rotating S0s, which show direct evidence of a
massive infall of gas and stars, also follow these trends.
\end{abstract}


%---------------------------------------------------------------------------

\section{Introduction: The S0 Problem}

The S0 galaxy class was invented by Hubble (1936) to fill a transitional
gap between ellipticals and spirals in his original morphological sequence
(Hubble 1926).  He proposed it as a way to ``include objects later than E7
but with no trace of spiral structure'' within his system, but in 1936 he
had no observational evidence for the class.  It was not until the 1950s
that definitive examples were identified from photographic surveys, and the
first systematic description of S0s in the literature was by Sandage (1961)
in {\it The Hubble Atlas of Galaxies}.

Why did it take so long?  One of the reasons is that in some sense S0
galaxies are classified not so much by what they are, as by what they are
not.  As a result, S0s are notoriously difficult to classify consistently
(e.g., Knapp et al. 1989).  Once you divide non-irregular galaxies between
ellipticals and spirals, S0s are what are leftover, and the fine details of
which category these belong in becomes strongly subject to the quality of
the observational material.  In selecting a sample for study, one
invariably encounters objects that on close inspection have no business
being on the observing list (see Eder 1990 and Pogge \& Eskridge 1993 for
examples of what can turn up in CCD imaging studies).

Despite these difficulties, a general picture of their basic properties has
emerged.  S0s are galaxies with massive bulges and weak disks without
spiral structure, and dominated by an old stellar population.  They have
ISMs comprised of a small quantity of cool gas (atomic and molecular) and a
small amount of dust, and can occasionally have a substantial hot ISM.  The
strongest departure from the traditional picture is the recognition that a
good fraction of all {\it bona-fide} S0 galaxies have a reasonably
substantial interstellar medium.

Of special concern to us here are the cool components of the ISM:
\hi, H$_2$ (CO), and dust (FIR).  From surveys of early-type galaxies
(Bregman et al. 1992; Hogg et al. 1993, and references therein), the state
of our knowledge of the cool ISM in S0s may be summarized as follows:
\begin{description}
\item[\hi\ 21-cm:] Between 30--40\% of S0s are detected, with
     \hi\ masses in the range of $10^7\la M_{HI}\la 10^{10}~M_{\sun}$.
\item[H$_2$ (CO):] The $^{12}$CO detection rate for S0s is $\sim25$\%,
     with inferred H$_2$ masses also in the range of $10^7\la M_{H2}\la
     10^{10}~M_{\sun}$
\item[Dust (FIR):] The FIR (60 and 100\micron) detection rate for S0s
     is $\sim40-60\%$, implying dust masses in the range of $10^4\la
     M_{dust}\la 10^{7}~M_{\sun}$ and typical dust/gas ratios.
\end{description}
In general, the cool ISM properties of S0 galaxies lie intermediate between
those of Ellipticals and Spirals (Hogg et al. 1993).

While the ISMs of S0 galaxies are reduced in scale relative to spirals,
they seem to have the same basic components in roughly similar proportions.
This has led to the question of whether there is enough interstellar
material in the right places for S0s to form new stars in the present day.
This paper reviews what has been learned about on-going massive star
formation in S0 galaxies, and what can be learned about the ISM in these
galaxies from the properties of the \hii\ regions, especially the gas-phase
abundances.  We compare the star formation properties of S0s with normal
spirals, and conclude with some questions for further study.

%---------------------------------------------------------------------------

\section{Present-Day Star Formation in S0s}

The most direct way to ask whether S0 galaxies are forming stars now is to
search for H$\alpha$ emission from \hii\ regions using narrow-band imaging.
There have been two main studies to date that we will focus on: the Lick
Survey of Pogge \& Eskridge (1993, henceforth PE93), and the CTIO study of
Caldwell, Kennicutt, \& Schommer (1994, henceforth CKS).

PE93 (also Pogge \& Eskridge 1987) surveyed 40 gas-rich S0 galaxies using
on-band/off-band filter imaging techniques.  This work used a 1-m telescope
with 75\AA\ bandpass filters in redshifted H$\alpha$+[N~II] emission-line
bands and adjacent emission-free continuum bands.  Of the 40 galaxies in
their combined sample, 8 were rejected as mis-identified (most were
misclassified irregulars that look lenticular on POSS plates).

CKS studied 8 S0 galaxies using Fabry-Perot imaging on the CTIO 4-m
telescope.  The Fabry-Perot etalon had a very narrow bandpass (2.4\AA\
FWHM), delivering roughly 10-times the sensitivity to faint emission
against the bright galaxy background of the filter imaging used by PE93.
This allowed the detection of fainter \hii\ regions, and for much deeper
upper limits on non-detections.

Combining the results of these studies and taking account of interlopers
and overlaps, of the 40 S0 galaxies studied, 18 show clear evidence of
significant populations of disk \hii\ regions.  The \hii\ regions are
primarily distributed into thin rings (some appearing as tightly wound
spiral arms), with a range of galactocentric radii from kiloparsec-scale
``nuclear rings'' to ``outer'' rings of 10--20~kpc diameter.  The remaining
22 S0 galaxies have no disk \hii\ regions, with upper limits such that we
should have detected single \hii\ regions powered by a single O5 star if
present.  In CKS's F-P study, the limits were very low, such that they
could detect an \hii\ region as faint as the Orion Nebula if present.
Faint diffuse disk emission and nuclear emission-line regions are also seen
in some of the galaxies upper limits, but nothing identifiable as a
discrete \hii\ region.

\begin{figure}[t]
  \plotone{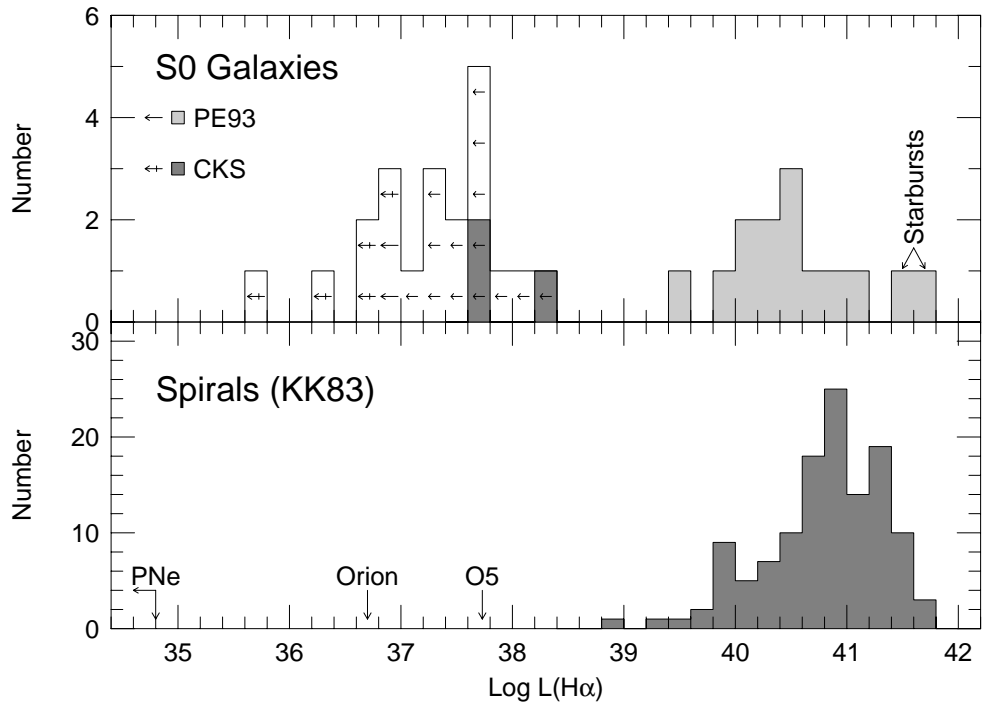}\\
  \caption{Integrated H$\alpha$ luminosities of S0s (top) and 
  Spirals (bottom).  Upper limits are shown as arrows.}\label{pogger:fig1}
\end{figure}

The distribution of integrated H$\alpha$ luminosities for these galaxies is
shown in Figure~\ref{pogger:fig1}, along with the upper limits for the
combined data sets (two of CKS's galaxies are in common with ours, and we
adopt their values).  The upper panel shows the S0s, while the lower panels
shows the data for Spiral galaxies studied by Kennicutt \& Kent (1983).
The luminosities of single \hii\ regions (one O5 star and Orion) are
indicated in the lower panel, along with the luminosity of the brightest
planetary nebulae (taken from CKS).  On average, the global star formation
intensity in these galaxies is lower than in spiral galaxies by about a
factor of 10, with the faintest extending to 2--3 orders of magnitude below
spirals.  To get some feel for how low it gets, the three S0 galaxies
detected by CKS each had $\sim$15 \hii\ regions, with individual regions as
faint or fainter than the Orion Nebula.

As CKS have correctly pointed out, in many cases our PE93 upper limits
would not have permitted us to detect some of the faintest \hii\ regions
detected by their F-P study.  Unfortunately, the only two galaxies we had
in common were non-detections for both studies, so we cannot assess how
much, if any, one should modify the non-detections reported by us.  Still,
even with these caveats in mind, it is clear that there exists a
significant population of S0 galaxies with no star formation down to very
faint limits.

Why the strong differences between star-forming and quiescent S0s?  One
interpretation (close to that discussed by CKS) is that star formation can
proceed all the way down to zero with no minimum cutoff.  What one is then
seeing in the S0 galaxy population is a steady fading of star formation
from a previous active state, active in this case being at the level seen
in Sa galaxies.  Indeed, during its ``active'' state, the galaxy would
probably be classified as Sa instead of S0.  Another possibility (discussed
by PE93) is that the appearance of the \hii\ regions in rings is strongly
suggestive that the same kinds of dynamical threshold effects that mediate
the rate of global star formation in spirals (Kennicutt 1988) is operating
in the S0s.  The S0s straddle the threshold for triggering star formation,
and so galaxies with similar general properties and ISM masses (on average)
can have radically different star formation properties.

At issue is whether there is a lower-limit to the star formation rate in
S0s or whether it just fades away.  A central problem is that neither
observational study is on firm statistical grounds.  Indeed, both studies
have made only a detailed reconnaissance of the problem to establish both
the commonality of star formation in S0s and the lower limits of its
intensity.  What is needed is a study undertaken with a carefully selected
sample of field S0s (to avoid the considerable influence of the hot
intra-cluster environment), and with careful attention to the problems of
assessing the detection limits.

%---------------------------------------------------------------------------

\section{\hii\ Region Abundances}

Having established that at least some S0s have \hii\ regions, what else can
be learned from the \hii\ Regions themselves?

\begin{figure}[ht]
  \plotone{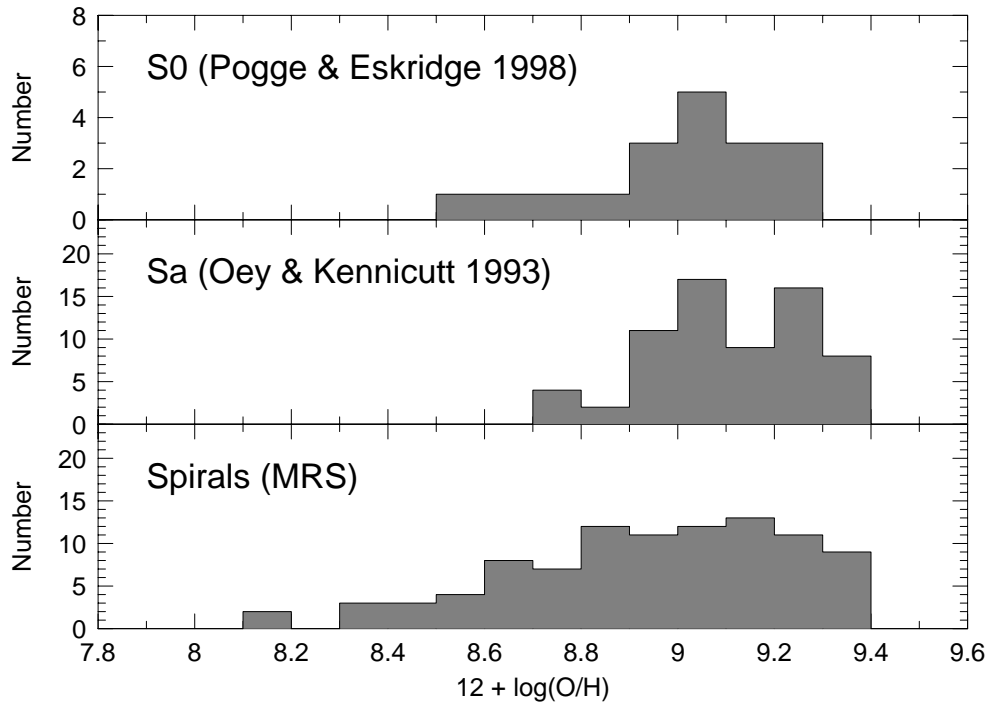}\\
  \caption{Distribution of 12+$\log$(O/H) for \hii\ regions in S0, Sa, 
           and late-type Spiral Galaxies.}\label{pogger:fig2}
\end{figure}

%The other line of inquiry is to obtain spectroscopy of the \hii\ regions
%with the intention of measuring the gas-phase abundances (especially the
%well-calibrated diagnostics for measuring O/H).  Such studies are limited
%to the brightest \hii\ regions in these galaxies, as you need high
%signal-to-noise ratio spectra to make these measurements reliably.  This is
%a project we have been pursuing with the brighter examples of \hii\ regions
%in our sample of star-forming S0 galaxies.

The gas-phase abundances of the elements derived from \hii\ regions provide
essential data on the origin and evolution of the ISM in S0s.  If the ISM
is derived primarily by accretion (infall) of new gas, then with typical
hydrogen (\hi+H$_2$) masses of order $10^{8-9}~M_{\sun}$, the most likely
donors are gas-rich dwarfs or small intergalactic gas clouds.  The
abundances of such objects are observed to be significantly sub-solar
(e.g., Skillman et al. 1989).  Even if the infalling material were adding
to a small amount of enriched intrinsic ISM gas shed by evolving stars, the
gas-phase abundances should still be significantly diluted compared to the
inferred stellar abundances.

On the other hand, if S0s are capable of replenishing their ISMs via mass
loss from evolved stars (AGB stars and PNe) and SNIa, one expects solar or
greater gas-phase abundances as is seen in the donor stars.  In a classic
computation, Faber \& Gallagher (1976) argued that a typical S0 or E galaxy
could build a substantial ISM in the mass ranges we are seeing in a few
billion years.

To date, we have obtained spectra of sufficient quality to estimate
extinctions, densities, and gas-phase oxygen abundances in 16 \hii\ regions
in six S0 galaxies.  Because none of the \hii\ regions had detectable
[O\,III]$\lambda$4363\AA\ emission, we estimated O/H abundances in these
\hii\ regions using the $R_{23}$ method of Edmunds \& Pagel (1984) and
adopting the calibration of Dopita \& Evans (1986) following Oey \&
Kennicutt (1993).  We find that all of the \hii\ regions have roughly solar
O/H abundances, with a range between 0.8 and 2 solar and a median O/H of
1.4 solar, adopting the solar value of $12+\log(O/H)=8.93$ (Anders \&
Grevesse 1989).  These values have a systematic uncertainty of 0.2~dex for
the absolute calibration (e.g., Oey \& Kennicutt 1993).

Figure~\ref{pogger:fig2} shows a comparison of our S0 galaxy \hii\ regions
with a sample of Sa galaxy \hii\ regions observed by Oey \& Kennicutt
(1993) and \hii\ regions in late-type spirals by McCall, Rybski, \& Shields
(1985).  The S0 galaxy \hii\ regions most resemble the Sa galaxies in their
abundances, and on average reside in the metal-rich part of the range seen
in late-type spirals.  In general, the metallicities of the S0 galaxy \hii\
regions do not resemble those of gas-rich dwarf irregulars, which range
from a few percent of solar up to about 0.5 solar for the most massive
(Skillman et al. 1989).

%---------------------------------------------------------------------------

\section{S0s and the Hubble Sequence}

The star formation and \hii\ region properties of S0 galaxies appear
remarkably normal when placed among the rest of the sequence of spiral
galaxies.  As noted above, S0 galaxies appear to continue the general trend
of decreasing star-formation rate with earlier Hubble Type observed by
Kennicutt \& Kent (1983), extending to the lowest instantaneous SFRs
measured (CKS).

\begin{figure}[ht]
  \plotone{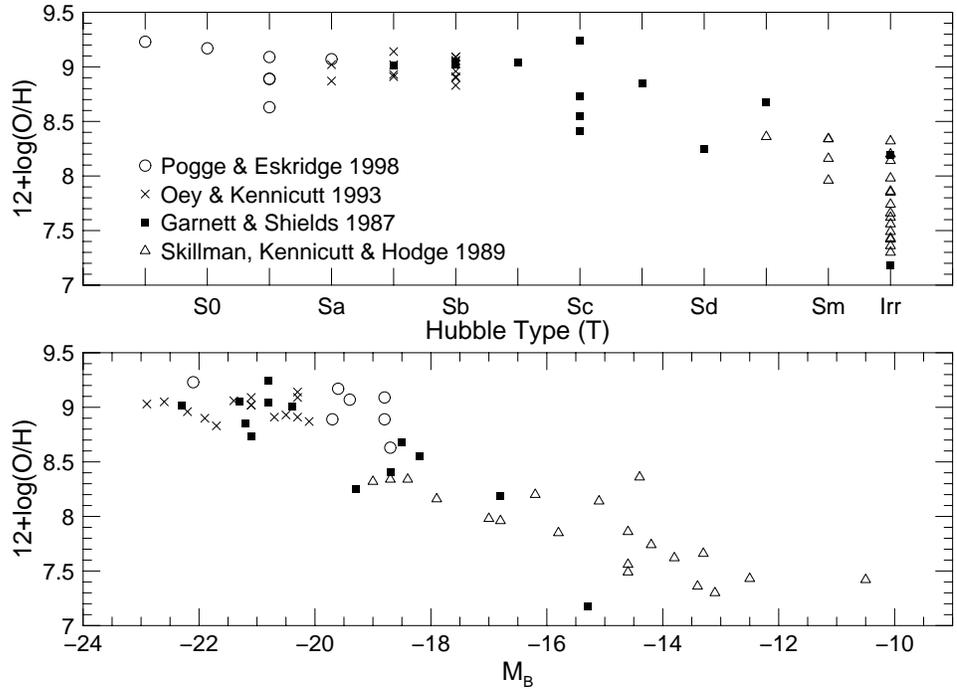}\\
  \caption{Correlation of 12+$\log$(O/H) with Hubble Type (top) and
           and total Luminosity (bottom) for S0 through Irr Galaxies.}
  \label{pogger:fig3}
\end{figure}

Among spirals and irregular, there is a clear trend of decreasing oxygen
abundance (O/H) with both later Hubble Type and decreasing Luminosity
(e.g., Roberts \& Haynes 1994; Garnett \& Shields 1987; Skillman,
Kennicutt, \& Hodge 1989).  If we add our S0 galaxy results to the
correlation diagrams plotted by Roberts \& Haynes (1994) for Irr through Sa
galaxies (Figure~\ref{pogger:fig3}), we find that the S0s fit onto the
observed trends.  Thus the metallicities that we observe for S0 galaxy
\hii\ regions are typical for galaxies of their luminosity, and continue
the observed trend with Hubble Type.  If the ISM in S0s was primarily
external in origin, due to recent infall of gas-rich dwarfs, we would have
expected to see systematically lower O/H abundances than predicted from the
trends for the galaxy luminosities, with a greater overall dispersion.  The
data therefore suggest that S0s are the lower end of the Hubble Sequence of
Spirals, without having to invoke recent infall to explain their
interstellar gas or star formation properties.

%---------------------------------------------------------------------------

\section{Counter-rotating Disks: A Conundrum?}

If only it were so simple.  Two of the six galaxies discussed in the
previous section, NGC\,4138 and NGC\,3593 are examples of S0/Sa galaxies
with counter-rotating disks of stars and gas (NGC\,4138, Jore et al. 1996;
NGC\,3593, Bertola et al. 1996).  In both, the \hii\ region rings are
apparently counter-rotating.  The model for the formation of such
structures, by the slow retrograde accretion of a gas-rich dwarf galaxy
(e.g., Thakar \& Ryden 1996; Thakar et al. 1997), would suggest that the
metallicities should be as low as in the putative donors.  Instead, we find
metallicities of 1.7 and 1.5 times solar in NGC\,4138 and NGC\,3593,
respectively.

\begin{figure}[ht]
  \plotone{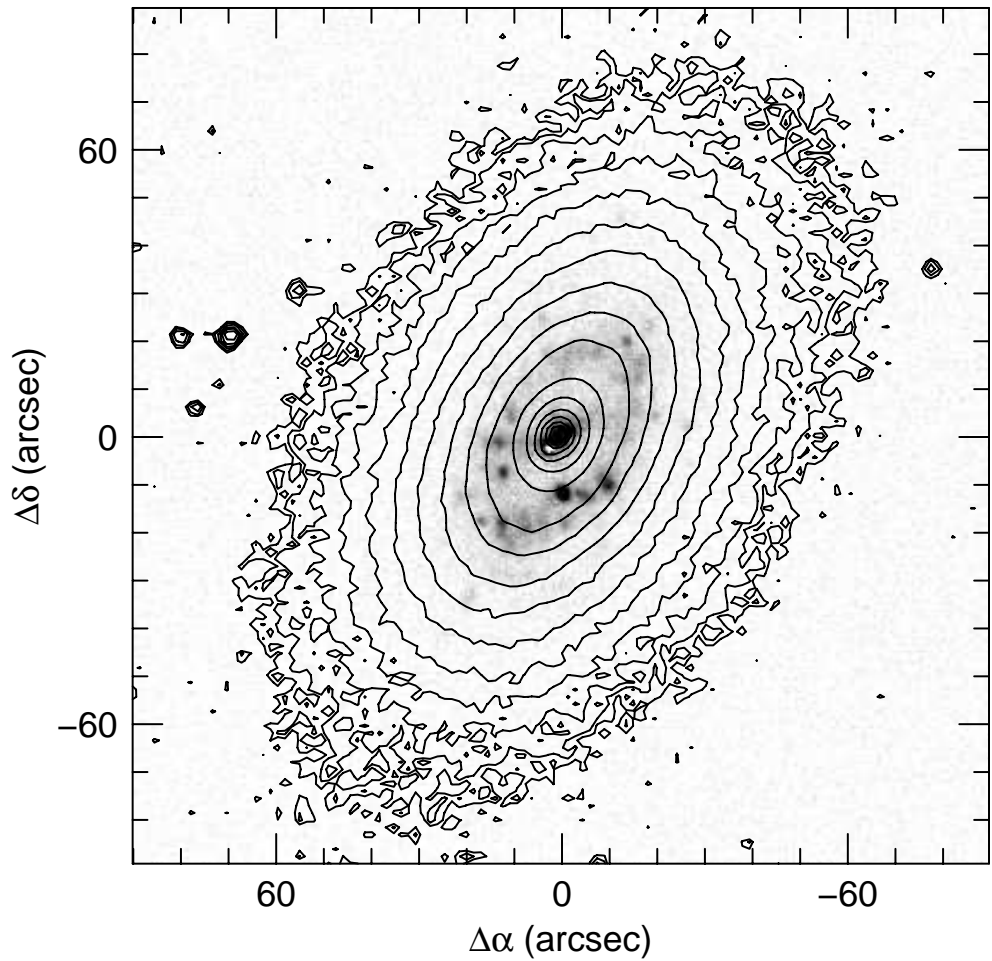}\\
  \caption{H$\alpha$+[N\,II] emission image of NGC\,4138 with
  H-band (1.6$\mu$m) contours superimposed.}\label{pogger:fig4}
\end{figure}

Further, both galaxies are morphologically undisturbed.  The ring of \hii\
regions in NGC\,4138 is perfectly aligned with the stellar light in visible
and near-IR images (Figure~\ref{pogger:fig4}).  The same is true of
NGC\,3593 (Corsini et al. 1998, also Pizzella et al., these proceedings).
There is nothing in their outward appearance to suggest they are the sites
of past minor mergers, even though there is strong dynamical evidence of
just such events.  What is going on?  Why are these structures both
apparently relaxed {\it and} metal rich?

One possibility is that the infall/merger event that gave rise to the
counter-rotating component occurred a long time ago, and formed a structure
that can be stable for the lifetime of the galaxy.  This would give it
enough time to self-enrich the gas compared to the original abundances of
the donor.  Dynamical evidence of a merger, however minor, does not
necessarily imply that it happened yesterday (i.e., that such structures
are ``smoking guns''; the gun may have cooled and the smoke cleared a long
time ago).  In particular, current models of hierarchical galaxy formation
suggest that merging among galaxies was much more common in the past, and
it is plausible that what we are seeing today is but a relict of the events
that made these galaxies S0s (e.g., Bekki 1998).  The fact that these two
galaxies are morphologically undisturbed and otherwise non-descript argues
that we need to look more closely at the longevity issue. 

%---------------------------------------------------------------------------

\section{Summary and Future Directions}

The star formation properties of S0 galaxies may be summarized as
follows:
\begin{itemize}
   \parsep 0ex
   \item A significant population of S0 galaxies is actively forming stars,
         while a similar population is quiescent down to the lowest
         measurable levels.

   \item In the star-forming S0s, the star formation rate is on average
         10--100 times lower than in spirals, and, in a couple of systems,
         it is as much as $\sim1000$ times lower.

   \item The \hii\ regions have solar or greater gas-phase oxygen
         abundances that are well within the normal range for spiral
         galaxies of similar luminosity.
\end{itemize}
In general, the trends of global star-formation rate and O/H abundances are
consistent with S0 galaxies populating the low end of the Hubble Sequence
of Spirals; there is no obvious discontinuity of properties from Sd through
S0.  Whatever the origin and subsequent evolution of the ISM is S0s (at
least in the field), it is the same as in spirals, only there is less of
it.  A number of interesting questions are raised by these results:
\begin{enumerate}

\item Is there a strong lower cutoff in star formation in S0s, or does star
      formation just fade away?  Is the apparent cutoff a function of the
      detection limits?  It is not simply a matter of gas content, since
      they are roughly similar in the star-forming and quiescent galaxies.
      Perhaps it is not how much gas a galaxy has so much as where it is
      and what form it is in (atomic or molecular) that matters.  Until
      there are systematic studies of statistically well-defined samples
      with good \hi\ and CO mapping (and not just reconnaissance), we will
      not be able to resolve this question.

\item How does the \hii\ region luminosity function for star-forming S0s 
      compare to that of later-type spirals?  This is essentially
      unexplored territory, but would go a long way towards learning
      whether the global star formation process is proceeding differently
      than in spirals (cf. the work of Caldwell et al. 1991 on Sa
      galaxies).  It is clear from the work surveying for \hii\ regions
      that the challenge is not just to find \hii\ regions, but to find
      enough of them faint enough to build a luminosity function.

\item Why are two S0 galaxies with counter-rotating gas and stellar disks, the
      most obvious examples of galaxies that have experience substantial
      infall of interstellar material, more metal rich by an order of
      magnitude than their most likely donor population?  Is this a fluke
      or a generic property of S0 galaxies with counter-rotating systems?
\end{enumerate}

Finally, what {\it are} S0 galaxies?  It has been suggested that if they
have detectable star formation, they are not S0s.  However, it is dangerous
to base the classification on parameters that are critically dependent on
the observation conditions (cf. CKS's discussion along these lines).  Star
formation isn't part of the classification criteria per se, although some
prominent features of spirals are a consequence of star formation activity
(e.g., the high contrast of spiral arms on blue photographic plates).  In
general, it is both unwise and contrary to the spirit of the {\it
morphological} classification system to attempt to impose secondary
physical criteria on any particular subclass.  In the end, the only honest
answer to the question of what are S0 galaxies is: we don't know, but it
looks as if we're on the road to starting to learn how to ask the right
questions that will lead to more satisfying answers.

%---------------------------------------------------------------------------

\acknowledgments

We wish to thank the conference organizers for hosting a fun and
stimulating meeting in a pleasant and relaxing atmosphere, and to the
participants for their lively and engaging discussions.  RWP's travel was
supported by the Department of Astronomy at The Ohio State University.

%---------------------------------------------------------------------------
%
% questions:
%

%---------------------------------------------------------------------------
%
% references
%

\end{document}